\documentclass[]{spie}  

\usepackage{amsmath,amsfonts,amssymb}
\usepackage{graphicx}
\usepackage[colorlinks=true, allcolors=blue]{hyperref}
\usepackage{enumitem}

\title{The Santa Cruz Extreme AO Lab (SEAL): Design and First Light}

\author[a,b]{Rebecca Jensen-Clem}
\author[b]{Daren Dillon}
\author[b,a]{Benjamin Gerard}
\author[a]{M.A.M. van Kooten}
\author[a]{J. Fowler}
\author[b]{Renate Kupke}
\author[b,c]{Sylvain Cetre}
\author[a]{Dominic Sanchez}
\author[b]{Phil Hinz}
\author[a]{Cesar Laguna}
\author[d]{David Doelman}
\author[d]{Frans Snik}
\affil[a]{Univ. of California, Santa Cruz (United States)}
\affil[b]{Univ. of California Observatories (United States)}
\affil[c]{W. M. Keck Observatory (United States)}
\affil[d]{Leiden Observatory, Leiden Univ. (Netherlands)}

\authorinfo{Further author information: (Send correspondence to R.~Jensen-Clem)\\E-mail: rjensenc@ucsc.edu}

\begin{document} 
\maketitle

\begin{abstract}
The Santa Cruz Extreme AO Lab (SEAL) is a new visible-wavelength testbed designed to advance the state of the art in wavefront control for high contrast imaging on large, segmented, ground-based telescopes. SEAL provides multiple options for simulating atmospheric turbulence, including rotating phase plates and a custom Meadowlark spatial light modulator that delivers phase offsets of up to $6 \pi$ at $635\,$nm. A 37-segment IrisAO deformable mirror (DM) simulates the W.~M.~Keck Observatory segmented primary mirror. The adaptive optics system consists of a woofer/tweeter deformable mirror system (a 97-actuator ALPAO DM and 1024-actuator Boston Micromachines MEMs DM, respectively), and four wavefront sensor arms: 1) a high-speed Shack-Hartmann WFS, 2) a reflective pyramid WFS, designed as a prototype for the ShaneAO system at Lick Observatory, 3) a vector-Zernike WFS, and 4) a Fast Atmospheric Self Coherent Camera Technique (FAST) demonstration arm, consisting of a custom focal plane mask and high-speed sCMOS detector. Finally, science arms preliminarily include a classical Lyot-style coronagraph as well as FAST (which doubles as a WFS and science camera). SEAL's real time control system is based on the Compute and Control for Adaptive optics (CACAO) package, and is designed to support the efficient transfer of software between SEAL and the Keck II AO system. In this paper, we present an overview of the design and first light performance of SEAL.
\end{abstract}


\section{INTRODUCTION}
\label{sec:intro}  

The last three decades have been marked by the discovery of over 4000 exoplanets orbiting nearby stars. Unraveling these planets' compositions, climates, and formation histories requires spectra of their thermal emission and reflected starlight. Direct imaging is the only path forward for obtaining high signal-to-noise, high-resolution spectra of such diverse exoplanet atmospheres, but today's instrumentation cannot yet support the direct imaging of mature, close-in worlds that indirect techniques have shown to be ubiquitous. The overarching challenge for high contrast imaging in the 2020s is therefore to improve our sensitivities to faint exoplanets at small separations in order to image and characterize the atmospheres of worlds whose ages, masses, and temperatures overlap with Solar System planets. The coming generation of 30-meter-class observatories are necessary but not sufficient for meeting this challenge, in that modern instrumentation could not yet support the direct imaging of rocky planets in the habitable zones of the nearest low-mass stars even with 30-m telescopes\cite{ess2017}. 

The key area for technology development is adaptive optics (AO): the performance of state-of-the-art high contrast imaging systems such as GPI\cite{2014PNAS..11112661M} and SPHERE\cite{2019AA...631A.155B} is limited by the quality of the wavefront correction delivered by their extreme AO systems\cite{2016SPIE.9909E..0VB,2017arXiv171005417M}. Directly imaging Solar System scale planets will require advanced wavefront sensor capabilities such as predictive wavefront control (van Kooten et al in these proceedings\cite{vankooten21}) and focal plane wavefront sensing and control (Gerard et al in these proceedings\cite{fast_seal}).

The complexity of both atmospheric turbulence and facility-class AO systems, however, means that accurately simulating the effects of such new AO technologies is challenging. The preferred solution -- on-sky testing -- is often limited by practical considerations such as AO bench and sky access at high-demand observatories whose primary focus is science. 

Our goal in developing the Santa Cruz Extreme AO Lab (SEAL) is to empower students and postdocs to develop next-generation AO technologies for exoplanet imaging in an accessible lab environment that is closely coupled to an on-sky AO system. To this end, SEAL includes a variety of turbulence simulation techniques and wavefront control options, bound together by a real time control system that is designed to support the efficient transfer of software between SEAL and the Keck II AO bench.  

In this paper, we describe SEAL's optical design (Section \ref{sec:optical_design}), our first steps towards imparting atmospheric-scale turbulence with a custom spatial light modulator (Section \ref{sec:turbulene}), a three-sided pyramid wavefront sensor (Section \ref{sec:pywfs}), a combined focal plane wavefront sensing and science arm (FAST; Section \ref{sec: fast_seal}), a traditional coronagraphic science arm (Section \ref{sec:coronagraphy}), our real time control system (Section \ref{sec: rtc}), and finally our next steps for the testbed (Section \ref{sec:conclusion}).

\section{Optical Design}
\label{sec:optical_design}
SEAL requires an on-axis, low-wavefront error optical design at visible wavelengths. The purpose of the design is to relay well-formed pupils between the simulated telescope aperture and each of the devices, and provide a relatively slow beam (f/38.5) at the coronagraphic focus. The design must also accommodate a suite of wavefront sensors to be used in closed loop, and so located after the deformable mirrors. The optical path therefore consists of a series of sequential 4-f relays, producing properly-sized pupil images in each collimated space. The pupil diameters at each of the apertures and devices are listed in Table \ref{tab:pupils}.

\begin{table}[ht]
\caption{\textbf{Summary table of aperture and device shapes and sizes.}}
\label{tab:pupils}
\begin{center}
\begin{tabular}{|l|r|r|}
\hline
Pupil location & Size (mm) and shape & Pupil diameter (mm)\\
\hline\hline
Phase Plate & 20.6 (round) & 20.6\\
Meadowlark Spatial Light Modulator & 17.6 x 10.7 (rectangular) & 10.5\\
Keck Aperture & 22.6 (hexagonal, central obs) & 22.6\\
Iris AO segmented DM & 3.5 (hexagonal) & 3.45\\
Boston Micromachines Kilo-DM & 10.88 (square) & 9.66\\
ALPAO DM97 & 15 (round) & 13.5\\
\hline
\end{tabular}
\end{center}
\end{table}
All apertures, turbulence plates and DMs listed in Table \ref{tab:pupils} are conjugate to the same altitude and all deformable mirrors are used at an incident angle of less than or equal to $10^{\circ}$.  
\begin{figure}[h!]
\vskip -0.1in
        \includegraphics[width=\textwidth]{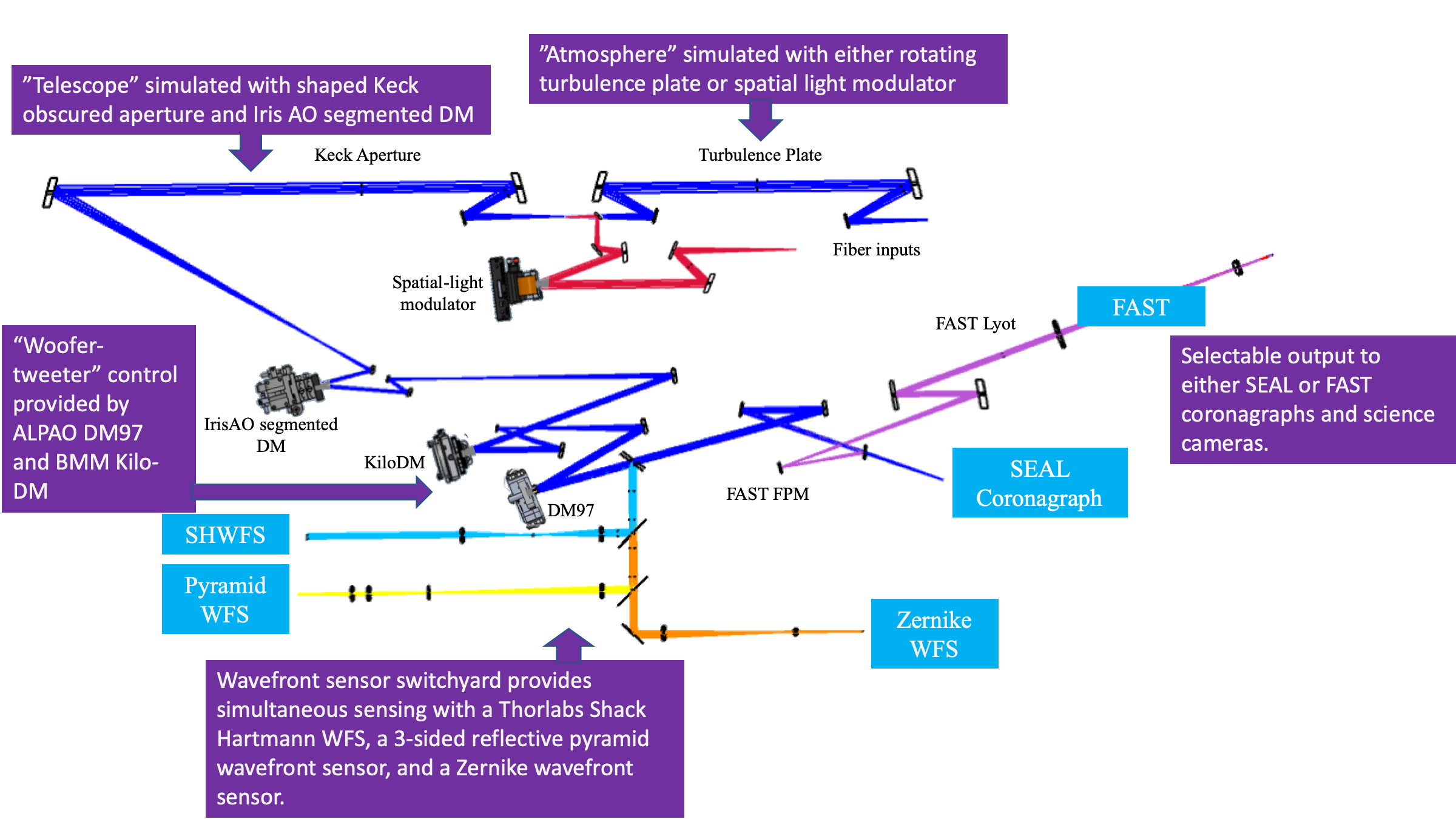}
        \caption{SEAL optical layout. The optical path starts at the top left fiber inputs, with either the rotating phase plate or SLM atmospheric simulation. Light continues to the left and down, encountering the telescope subsystem consisting of an obscured hexagonal aperture and segmented IrisAO DM. After the woofer/tweeter relays, some of the light is split (down in figure) to a switchyard with its selection of multiple wavefront sensors. The remaining light continues through to the right to the coronagraphic focus or FAST.}
        \label{fig:optical_layout}
\end{figure}
The full SEAL optical layout is shown in Figure {\ref{fig:optical_layout}}. The optical path can be thought of as a series of subsystems: atmosphere, telescope, deformable mirrors, wavefront sensors and coronagraph. In the atmospheric subsystem, a deployable fold allows the selection of simulated turbulence via a rotating phase plate or the Meadowlark spatial light modulator. The telescope subsystem contains both a centrally-obscured hexagonal aperture and the segmented IrisAO 37 actuator mirror which mimics the segmented Keck primary mirror.  Considerations for both the size of the phase plate and Keck apertures included manufacturability of turbulence plates with $D/r_0$ equal to 40, 60 and 66.7, as well as the secondary support spiders in the miniature Keck aperture. Wavefront control is provided by ``tweeter" (BMC KiloDM) and ``woofer" (Alpao DM97) deformable mirrors. In the collimated space after the ALPAO DM97 is a 50/50 beamsplitter which reflects light to the wavefront sensor switchyard. There we have beamsplitters to allow simultaneous operation of an off-the-shelf Shack-Hartmann wavefront sensor, a 3-sided reflective pyramid wavefront sensor and a vector Zernike wavefront sensor. The liquid crystal vector Zernike waveplate is identical to the device described in Doelman et al.~2019\cite{Doelman19}; the layout of our Zernike wavefront sensor arm has not yet been designed, but will follow the layout described in Doelman et al.~2019\cite{Doelman19} (including polarizing optics, a quarter waveplate, and a Wollaston prism to allow for the simultaneous sensing of phase and amplitude errors). Following the beam transmitted through the wavefront sensor dichroic, a final focusing optic provides an f/38.5 beam for both FAST and the traditional coronagraphic arm. \par The optical layout shown in Figure \ref{fig:optical_layout} uses off-axis parabolic relays to provide nearly perfect on-axis performance. The high precision ($\lambda/20$) optics for this design are currently being fabricated. Presently the testbed is operating with a fully refractive, achromatic lens-based system to allow for integration, optimization and debugging of the many devices utilized by SEAL. Upon completion of the off-axis parabolas the system will migrate to the reflective relay design shown in Figure \ref{fig:optical_layout}.

\section{Generating Turbulence with a Spatial Light Modulator}
\label{sec:turbulene}

\begin{figure} [ht]
   \begin{center}
   \begin{tabular}{c} 
   \includegraphics[width=.25\textwidth]{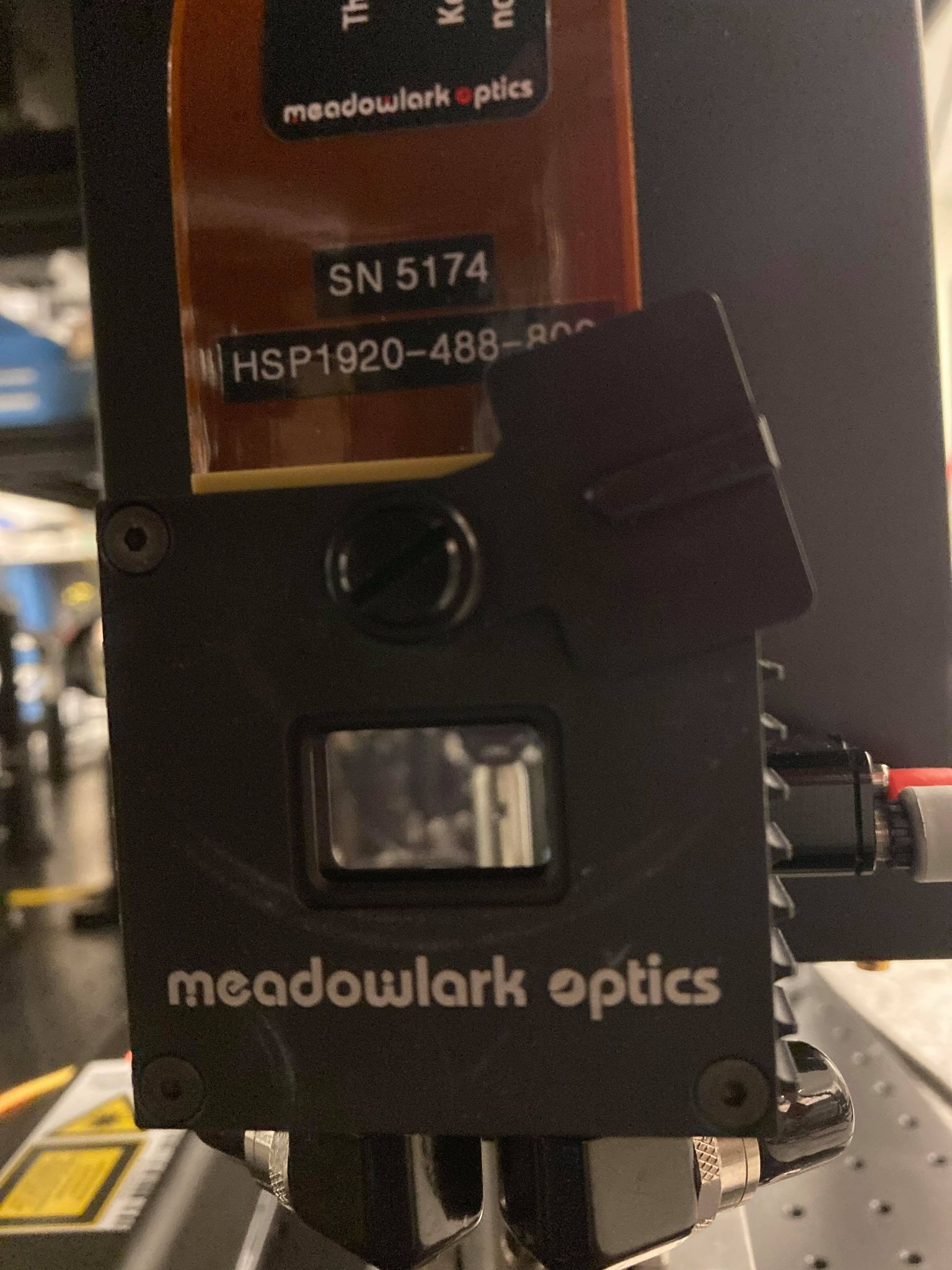}
      \includegraphics[width=.75\textwidth]{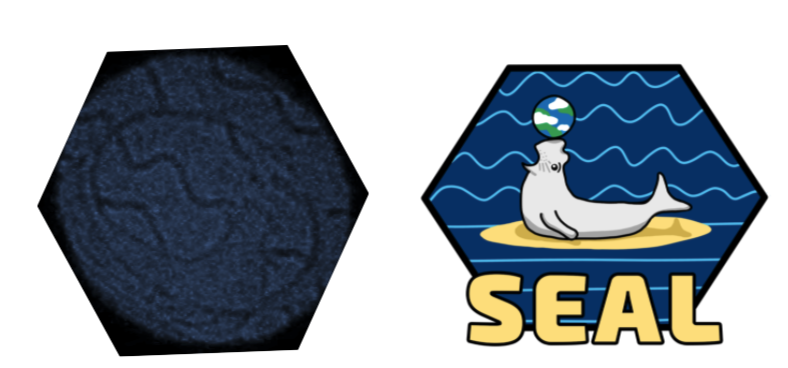}
	\end{tabular}
	\end{center}
   \caption[example] 
   { \label{fig:SLM_pics} 
The spatial light modulator (SLM) in our setup. Left: SLM viewed face on. Right: An example applied pattern (the SEAL logo) as reflected on a notecard to test spatial resolution finer than the Shack Hartmann wavefront sensor in our setup can measure.}
   \end{figure}

In order to introduce realistic turbulence in a laboratory environment, we are testing and integrating a custom reflective spatial light modulator (SLM) from Meadowlark Optics. An SLM uses liquid crystal pixels to apply phase patterns. The advantage of introducing turbulence with an SLM instead of a DM is in the spatial sampling of the turbulence: our SLM has 1920 by 1152 pixels, to make up a total area of 17.6 by 10.7mm \cite{SLMManual}, with a pixel pitch of 9.2$\mu$m and a 95.7\% fill factor. We acquired a custom SLM with a $6\pi$ dynamic range instead of the standard $2\pi$ in order to reproduce the typical variance across the wavefront at Keck. Noll 1976\cite{1976JOSA...66..207N} gives the piston, tip, and tilt subtracted variance as $\sigma^2=0.134(D/r_0)^{5/3}$. For $D=10\,$m and Mauna Kea's $20\%$ best $r_0=23.6\,$cm (KAON303\cite{Neyman04}), $\sigma=0.66\mu$m. Our SLM ``stroke'' of $1.9\mu$m at $\lambda=0.635\mu$m is sufficient to match up to $\sim3\times$ this value. The SLM has a response time of $422.4\,$Hz, so we will apply the turbulence and the AO correction somewhat slower than on-sky systems. Figure \ref{fig:SLM_pics} shows a photo of the SLM and a demonstration of an applied SEAL logo pattern.

\subsection{Ongoing Efforts to Calibrate the Spatial Light Modulator}

Before integrating the SLM into our extreme AO testbed, we aim to characterize the SLM both in space (pixel resolution due to any tied pixels or crosstalk and the max amplitude of the phase pattern we can measure from the device) and time (how quickly we can apply patterns). We plan to characterize any non-linearities in the response as well as make a flat map. Due to the small (9.2 by 9.2$\mu$m) pixel size of the SLM, the high amplitude of the max phase pattern, and the custum nature of the device, these calibration and flattening steps are nontrivial. 

\begin{figure} [ht]
   \begin{center}
   \begin{tabular}{c} 
   \includegraphics[width=.5\textwidth]{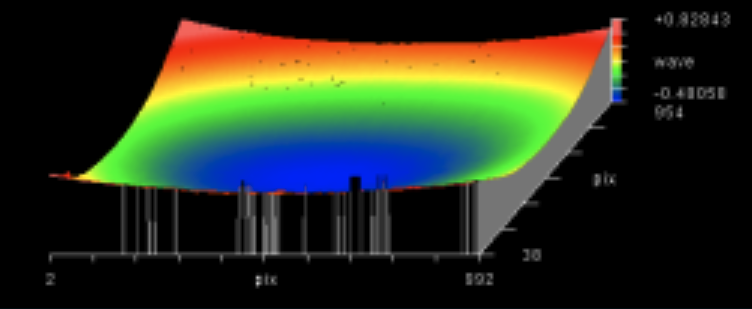}
   \includegraphics[width=.5\textwidth]{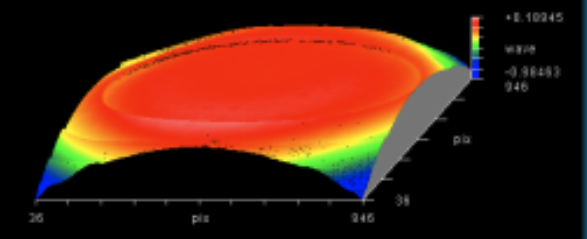}
	\end{tabular}
	\end{center}
   \caption{Zygo interferometer images of the Meadowlark SLM. Left: the natural shape of the SLM with zeros applied. Right: the residual power after applying the shipped flat. The scale is shown in waves for a 635nm laser source, with 156nm RMS on the natural shape and 96nm RMS after applying the Meadowlark flat.}
    \label{fig:zygo} 
   \end{figure}

Our first experiments started with placing the SLM in front of a Zygo Inteferometer. However an interferometer can only measure 2$\pi$ of phase, and relies upon reconstruction and phase wrapping to measure $>$2$\pi$. We found that the natural shape of the SLM was larger than the $2\pi$ the Zygo could recover, and we were unable to determine if unusual readings from the instrument were indicative of non-linearities from pixel pokes or the phase-wrapping reconstruction algorithm from the Zygo. While Meadowlark did ship the SLM with a factory flat, with this flat applied we still found what appeared to be more than $2\pi$ of phase pattern, most notably in a large defocus pattern imparting on the order of $\sim$100nm RMS (Figure \ref{fig:zygo}). Even with a beam expander in our setup, we could not resolve individual pixels, but rather we could only sample down to $\sim$5 SLM pixels. 

\begin{figure} [ht]
   \begin{center}
   \begin{tabular}{c} 
   \includegraphics[width=\textwidth]{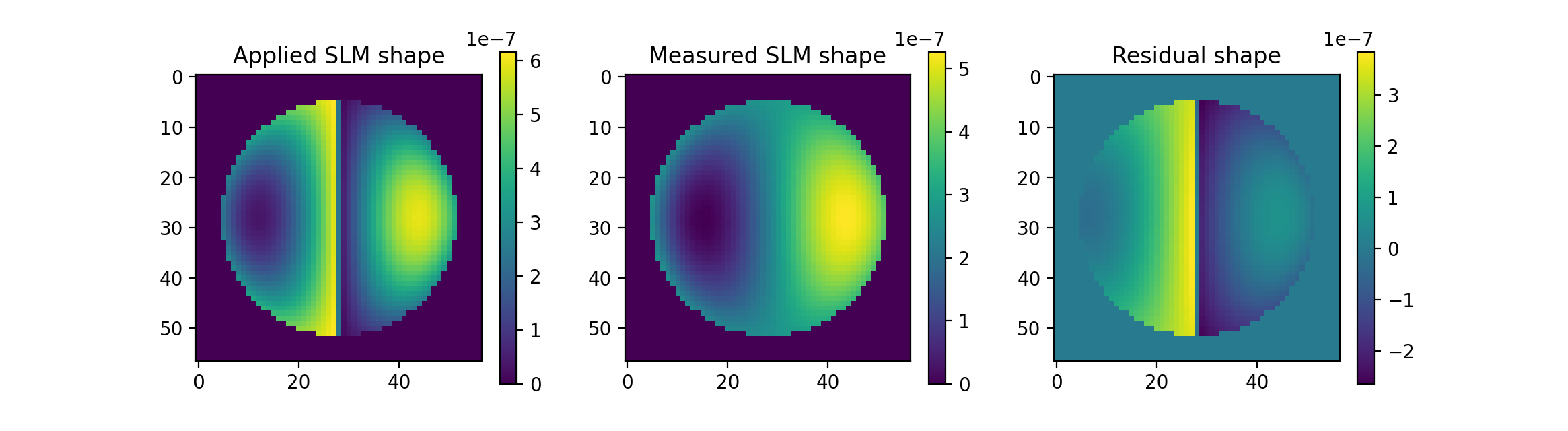}
	\end{tabular}
	\end{center}
   \caption{From left to right: the applied SLM shape, the measured aberration using the high order Shack-Hartmann wavefront sensor (35x35), and the residual when subtracting the applied and measured shapes. The colorbars indicate the wavefront error in meters. The assumed dynamic range of the SLM is taken to be $2\pi$ due to the miscalibrated linearization file. The mismatch in the center is due to the sudden transition in the SLM shape that cannot be accurately reconstructed by the wavefront sensor due to its low spatial resolution compared to the SLM. }
    \label{fig:zernike_basis} 
   \end{figure}

Next, we worked with a high order Shack-Hartmann wavefront sensor (SHWFS) from Thorlabs (model WFS-40), which lacks the resolution to sample more finely than a few hundred of the SLM's pixels but can reconstruct more than 2$\pi$ of applied phase. With the SHWFS we confirmed that the linear response of the SLM device was not as expected. We concluded that this was due to the linearization files that were shipped with the device -- these files had been calibrated and set to a max amplitude of the standard $2\pi$ rather than our custom $6\pi$ value. Figure \ref{fig:zernike_basis} shows our initial work to create a Zernike modal basis for our SLM, currently on a $2\pi$ scale.


At present we are working to build a calibration setup within the SEAL testbed to build our own set of linearization files that allow control of the device up to $6\pi$. Because this calibration (and therefore linear response of the device) is highly dependent on a given optical setup, our new setup will allow for recalibration of the SLM when we change our optical path by adding a beam splitter after the SLM allowing for an additional imager.

\section{Pyramid Wavefront Sensing}
\label{sec:pywfs}
A three-sided reflective pyramid wavefront sensor (3RPWFS) is currently being tested in SEAL. This pyramid sensor configuration eliminates chromatic aberrations, is easy to fabricate, and distributes the incoming light between three pupils rather than four in order to support slightly fainter targets for wavefront measurements. The SEAL 3RPWFS is a prototype design that will later be tested on-sky using the Shane Telescope on Mount Hamilton and therefore the design is motivated by that telescope's adaptive optics specifications. A more detailed overview of the optical design and motivation can be found in Sanchez et al.~2020\cite{sanchez_design_2020}.

\subsection{Optical Design and Lab Setup}
\begin{figure}[ht]
    \centering
    \includegraphics[width=\columnwidth]{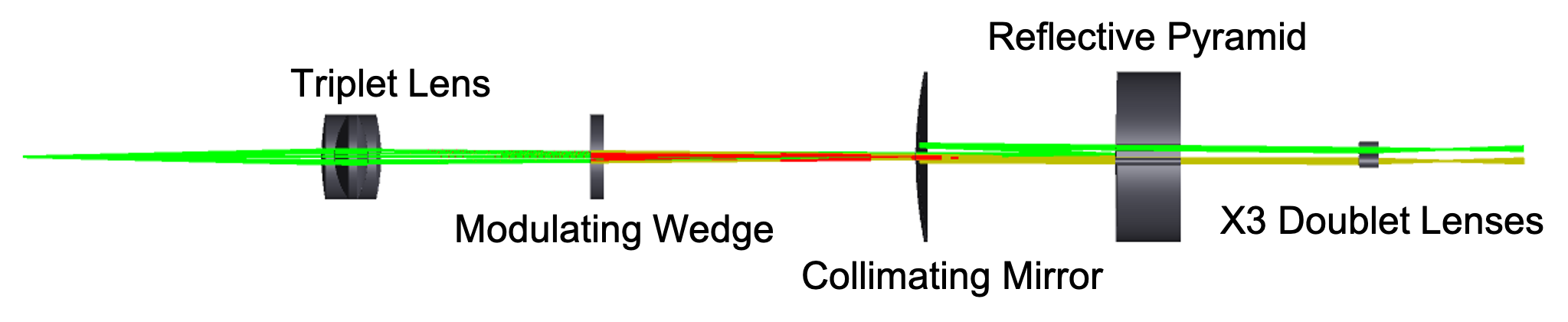}
    \caption{3RPWFS SEAL Optical Layout}
    \label{fig:3RPWFS_layout}
\end{figure}

The 3RPWFS in the SEAL testbed is not entirely reflective as shown in Figure \ref{fig:3RPWFS_layout}. The Shane Telescope’s current AO track only allows for an undeviated optical axis design and has a track length of about 226.35~mm. Therefore, this on-axis 3RPWFS design is being developed to demonstrate its performance and feasibility with a 500-900nm wavelength range. 

Like the Shane AO module, an incoming telecentric F/21.08 beam is re-imaged to produce an F/50 beam using a triplet lens. Because there is no possibility for a modulating mirror inside the Shane Telescope AO module, we instead place a glass wedge in the re-imaged pupil plane to modulate the incoming beam. Circular modulation is achieved by spinning the glass wedge about the optical axis. It is important to note that the modulation radius is fixed, since ray deviation is achieved by the wedge angle. In order to change the modulation radius, a different wedge must be used.

\begin{figure}[ht]
    \centering
    \includegraphics[width=\columnwidth]{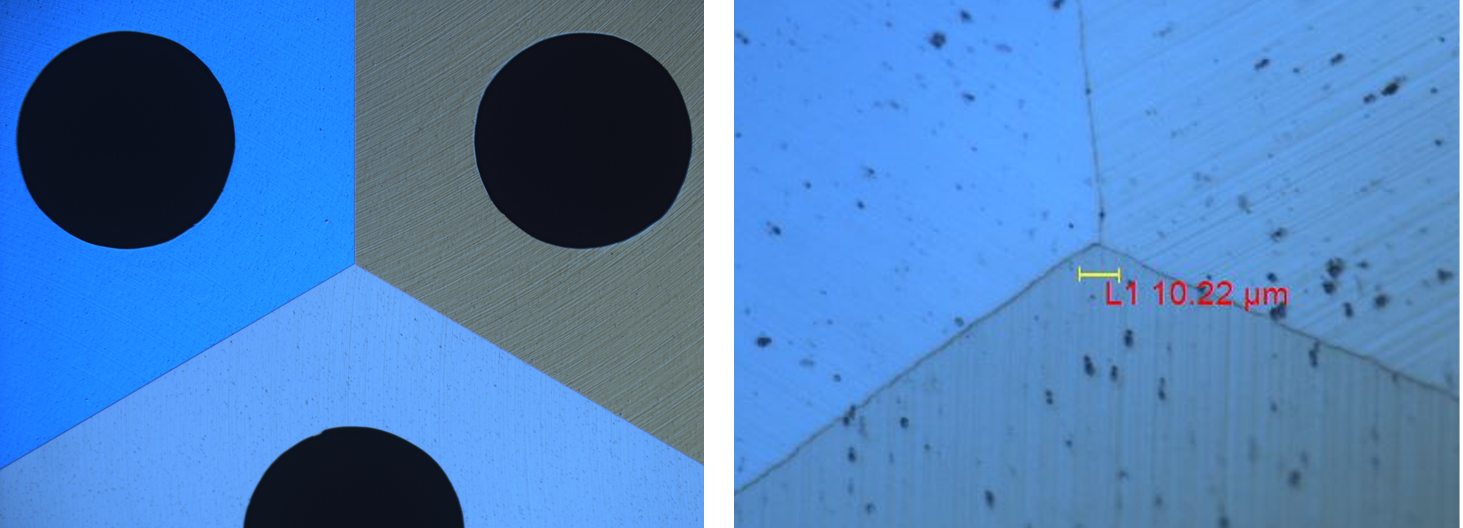}
    \caption{Left: a microscope image of the three-sided reflective pyramid illustrating the drill holes where the collimating pupils pass through. Right: a higher magnification image of the pyramid tip showing the roundness diameter of the pyramid apex.}
    \label{fig:3RPWFS_microscope}
\end{figure}

Upon reflection from the pyramid apex, a parabolic mirror with a drill hole at its vertex collimates the reflected light from the apex and re-images the pupils behind the pyramid. The reflective pyramid also has three drill holes --- one for each pupil --- to allow the pupil light through unvignetted. Images of the reflective pyramid are shown in Figure \ref{fig:3RPWFS_microscope}. The apex angle of the pyramid is constrained by the Shane AO detector. If the apex angle is too large, the pupil images will not fit within the detector; however, if the apex angle is too shallow, reflected light from the apex will not be collected by the collimating mirror and will be sent back through the collimating mirror drill hole. 

\begin{figure}[ht]
    \centering
    \includegraphics[width=\columnwidth]{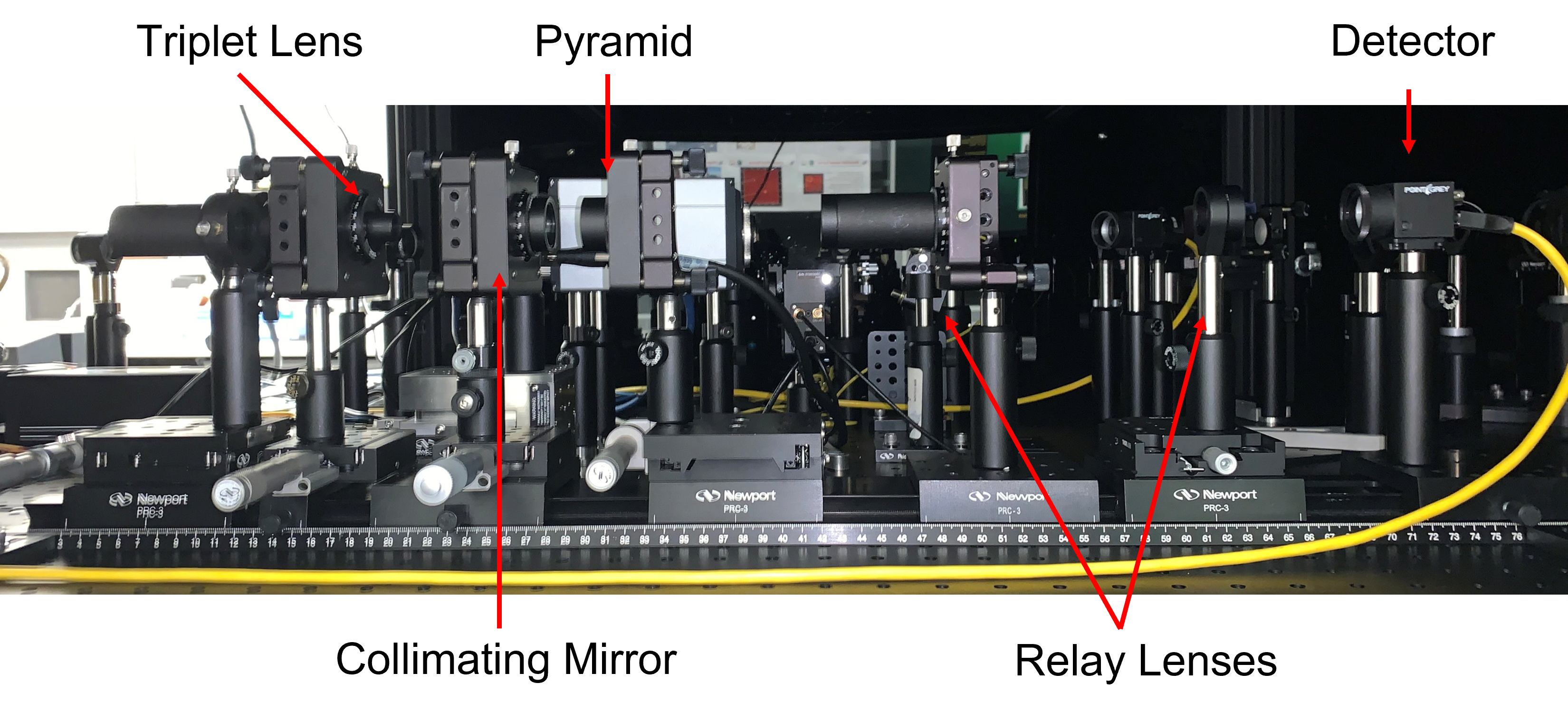}
    \caption{3RPWFS SEAL lab setup.}
    \label{fig:3RPWFS_labsetup}
\end{figure}

\begin{figure}[ht]
    \centering
    \includegraphics[width=6cm]{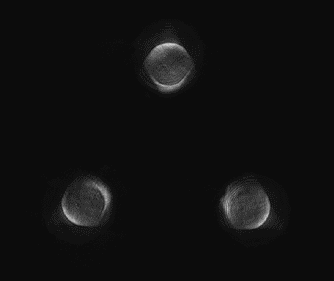}
    \caption{Preliminary pupil images of the 3RPWFS in SEAL testbed.}
    \label{fig:3RPWFS_pupilimages}
\end{figure}

The Shane AO detector is fixed and cannot be moved along the optical axis, which means that the re-imaged pupil images behind the pyramid must be re-imaged onto the detector. Given the limited space in Shane AO, three small doublet lenses will re-image their respective pupils onto the detector. The footprint of the pupils is expected to fit within the detector, with each pupil measuring $630\,\mu$m in diameter. The SEAL 3RPWFS is shown in Figure \ref{fig:3RPWFS_labsetup}; however, at the time of writing this paper, the three relay doublets have not yet arrived. Instead, two singlet lenses relay the pupil images onto the detector. The preliminary pupil images of the 3RPWFS are shown in Figure \ref{fig:3RPWFS_pupilimages}. These images were taken using a planar mirror to demonstrate the expected pupils because the ALPAO deformable mirror has yet to be flattened at the time of writing this paper.

\subsection{Preliminary Results}
\begin{figure}[ht]
    \centering
    \includegraphics[width=\columnwidth]{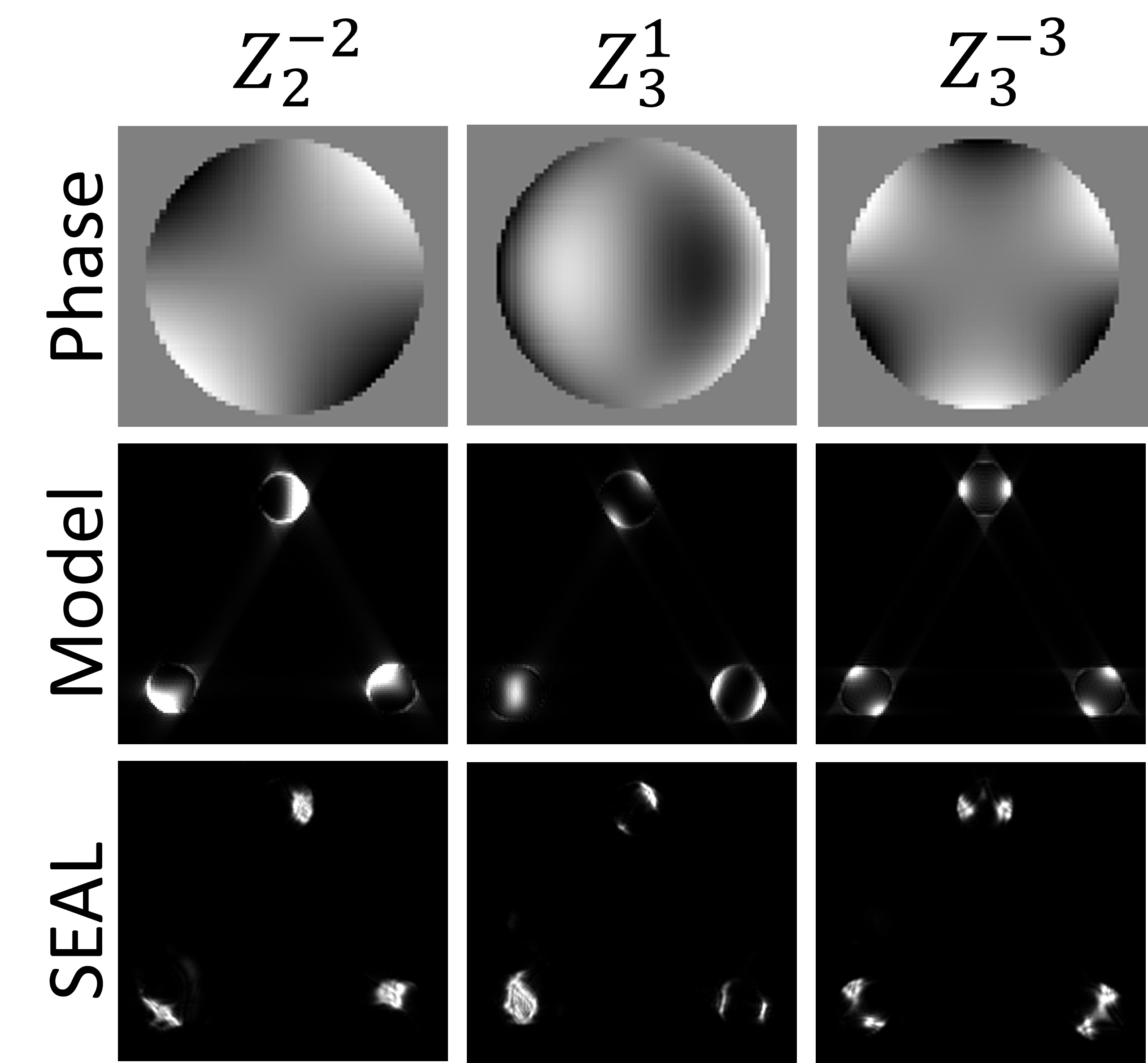}
    \caption{Top: HCIPy pupil images of induced Zernike polynomials onto the DM. Middle: the expected pupil images with the respective Zernike polynomial produced using HCIPy. Bottom: the pupil images produced by the SEAL 3RPWFS.}
    \label{fig:3x3-aberrations}
\end{figure}

The 3RPWFS pupil images are simulated using the python package High Contrast Imaging for Python (HCIPy\cite{por2018hcipy}) and are compared to the in-lab images. Figure \ref{fig:3x3-aberrations} shows a comparison between the HCIPy and testbed pupil images while inducing three different Zernike polynomials: $Z_{2}^{-2}$, $Z_{3}^{1}$, and $Z_{3}^{-3}$.  The top row of Figure \ref{fig:3x3-aberrations} illustrates the Zernike polynomials being applied to the DM; this is not to be confused as the actual DM surface, since the ALPAO DM cannot perfectly construct this shape. The middle row of Figure \ref{fig:3x3-aberrations} shows the expected 3RPWFS pupil images and is blind to the apex roundness diameter and other small imperfections in the pyramid. The bottom row of Figure \ref{fig:3x3-aberrations} shows the SEAL 3RPWFS pupil images for the respective aberrations. Although the ALPAO DM has yet to  be calibrated, the aberrated pupil images seem to generally confirm the behavior of the 3RPWFS.

\clearpage

\section{Focal Plane Wavefront Sensing}
\label{sec: fast_seal}
The SEAL testbed enables a high-speed focal plane wavefront sensing and control method known as the Fast Atmospheric Self-coherent camera Technique (FAST), described in detail in Gerard et al. (2021)\cite{fast_seal} from this same conference proceedings but also summarized in this section. FAST is built on the Self-Coherent Camera (SCC)\cite{scc_orig} focal plane wavefront sensing technique, which utilizes a custom Lyot stop design to form fringes in the coronagraphic image on remaining stellar speckles and diffraction but not on any off-axis exoplanet sources. These fringes enable wavefront sensing and reconstruction of speckle amplitude and phase in a single image (where normally without fringes you can only measure speckle amplitude, but not phase, in a single coronagraphic image), which in turn enables speckle correction/subtraction via real-time feedback to a DM and/or post-processing. FAST builds on the SCC by using a custom coronagraph focal plane mask, designed to increase the SCC image fringe visibility (i.e., the ratio of fringed to unfringed image components) by up to 6 orders of magnitude so that fringes can be detected on a speckle even when only tens of photons are recorded per speckle. As a result, this capability enables the SCC to measure and correct for residual atmospheric speckles in addition to quasi-static ones, where millisecond timescale exposures are needed to sufficiently measure and correct for the dynamically evolving atmosphere but for which few photons are collected in a coronagraphic image over that timescale.

\begin{figure}[h]
\centering
\includegraphics[width=1.0\textwidth]{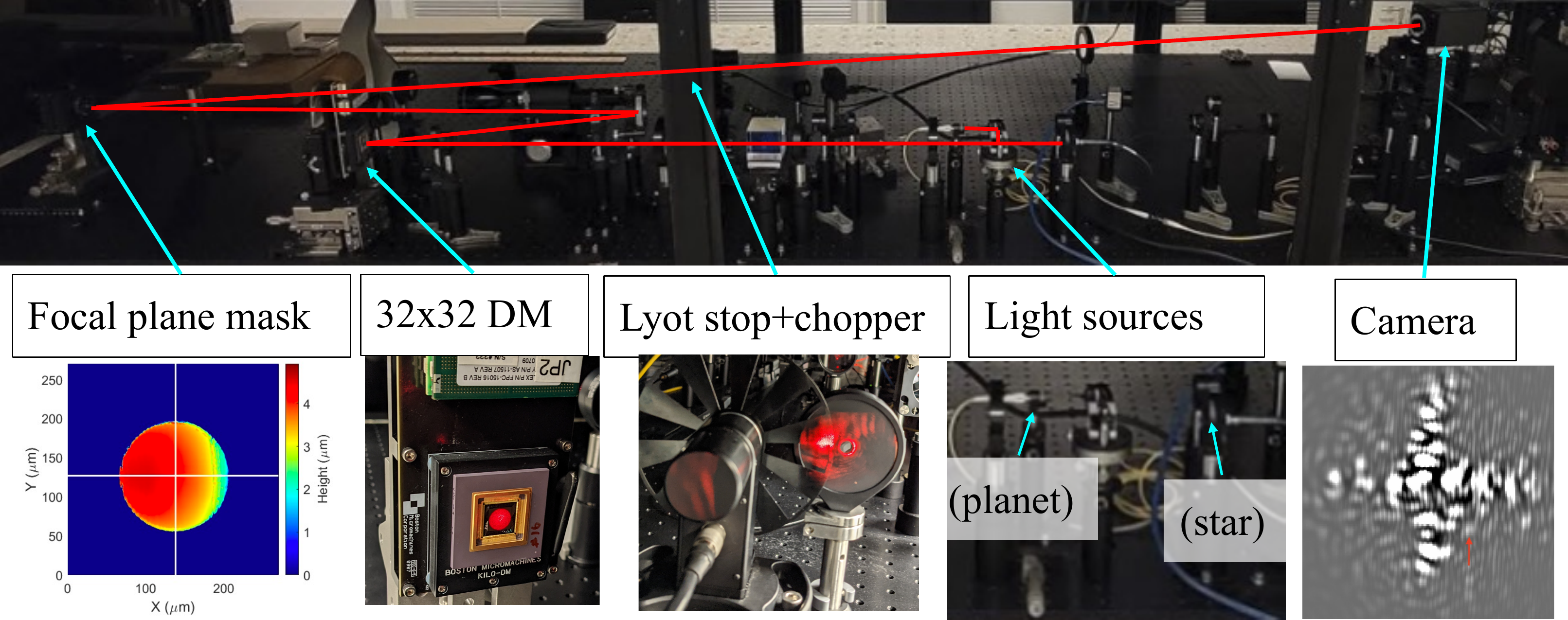}
\caption{An outline of our FAST SEAL setup, from Gerard et al. 2021 in these SPIE conference proceedings\cite{fast_seal}. A planet and star light source, both aligned to the same focal plane, are combined and then collimated, with the planet source positioned slightly off-axis a few diffraction bandwidths from the star. An aperture stop and re-imaging optics place the pupil on a 32x32 actuator DM, sampling 29 actuators across the beam diameter. The beam is then focused on our custom FAST focal plane mask, then re-collimated to a downstream coronagraphic pupil, where a Lyot stop and optical chopper modulate the beam. Finally, downstream imaging optics enable coronagraphic focal plane (i.e., FAST) or pupil plane imaging (the chopper and Lyot stop are on repeatable magnetic mounts to enable un-occulted coronagraphic pupil imaging.}
\label{fig: fast_seal}
\end{figure}

Our original and current refractive FAST setup was developed on the same granite testbed as SEAL, but using a separate set of light sources and optics. This setup has enabled FAST software development in parallel while the full SEAL testbed is being build and integrated. The main hardware components, illustrated in Figure \ref{fig: fast_seal}, include 
(a) two light sources to simulate a star and off-axis exoplanet (the latter with the option to add a neutral density filter at a user-defined optical density to choose what contrast ratio the planet should be set to), (b) a 32 x 32 actuator MEMS, (c) a custom FAST coronagraph mask made by the University of Alberta's nanoFAB laboratory, (d) a coronagraphic Lyot stop including an off-axis pinhole designed for the SCC, (e) a Thorlabs MC2000B optical chopper device to temporally modulate fringed and unfringed images, thereby enabling post-processing-based speckle subtraction called coherent differential imaging (CDI), and (f) an Andor Zyla 5.5 sCMOS camera. A Python 3 high-level software interface using low-level real-time control (RTC) software based on the Keck infrared Pyramid wavefront sensor RTC (Section \ref{sec: rtc}) enables an operational frame rate of 50 Hz (i.e., a $20\,$ms pause is needed in between acquiring images and sending DM commands). With this setup, in Gerard et al. (2021)\cite{fast_seal} we demonstrate real-time FAST DM correction of evolving AO residuals, showing a contrast gain of a factor 2-10, depending on separation. 

Several future tests are planned once FAST is integrated into the full SEAL testbed, including operation over larger spectral bandwidths (we are currently using a narrow band laser light source), optimized temporal control (including predictive control), and combining ``first stage'' real-time AO correction with ``second stage'' FAST. The latter will be enabled by our Thorlabs high-speed Shack Hartmann wavefront sensor (model WFS-20), which we have already benchmarked in our SEAL low-level RTC to enable 460 Hz frame rate correction with 26 subapertures across the wavefront diameter.

\section{Coronagraphy}
\label{sec:coronagraphy}
\subsection{Coronagraph Design and Optimization}

Here we describe plans for SEAL's traditional coronagraphic science arm. We simulated the performance of two coronagraphs: a classical Lyot coronagraph\cite{Sivaramakrishnan, Makidon} and a Vector Vortex coronagraph\cite{Mawet2009}, both through a continuous circular primary without a central obscuration or support spiders.
For each design, we used HCIPy\cite{por2018hcipy} to simulate perfect and realistic versions of the designs; realistic designs include wavefront error from imperfect optical surfaces and amplitude error from imperfect optical edges. With optics that might impart up to 10nm RMS of wavefront error added in quadrature over 10 optics in our system, we simulate $32$nm RMS with a power spectral density (PSD) of -2 to correspond to polished optics. For amplitude errors we allow for irregularity around the edge of the Lyot stop with a resolution of 30$\mu$m, which corresponds to $\sim$0.01 of the Lyot stop diameter (varying subtly by the Lyot stop diameter design.) 


As metrics to evaluate the performance of our designs we use contrast curves normalized to planet throughput and integration time to reach an SNR of 1, as described in Ruane et al. 2018 \cite{Ruane2018}. 
While an idealized vortex offers unparalleled light suppression and imaging at very small inner working angles, the vortex is also more susceptible to the effects of wavefront error. Similarly, vortex coronagraphs are more difficult to integrate and align for an initial testbed setup.

\subsection{Preliminary Designs and Hardware}

\begin{figure}[ht]
    \centering
        \includegraphics[width=.45\textwidth]{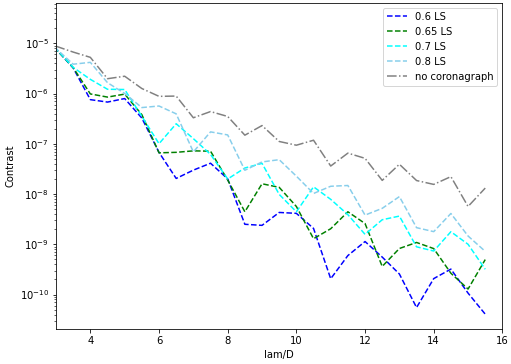}
    \includegraphics[width=.54\textwidth]{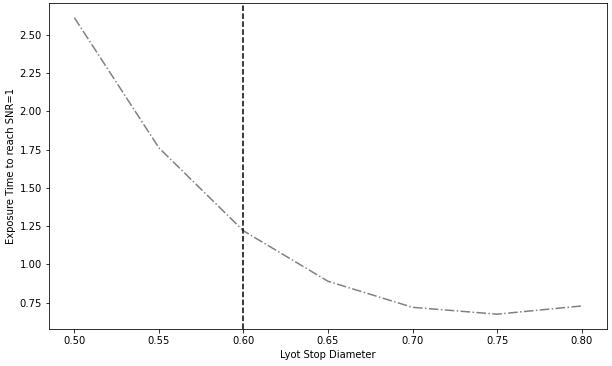}
    \caption{The design for the classical Lyot coronagraph. Our focal plane mask diameter is set at $\frac{3\lambda}{D}$, where D is the diameter of the primary mirror -- to allow for maximum light extinction without blocking close in companions. Left: the contrast curves. For a classical Lyot design, an increasingly undersized Lyot stop will create notably better contrast, but reduce the throughput from faint secondary sources. Right: the exposure time to reach SNR$=1$ for a planet observation. Lyot stops larger than $\sim$0.65 have a similar exposure time to reach $SNR=1$. To optimize both contrast and exposure time, we target an optimized design with a Lyot stop diameter of 0.6 relative to the primary diameter. }
    \label{fig:lyot_design}
\end{figure}

\begin{figure}[ht]
    \centering
        \includegraphics[width=.45\textwidth]{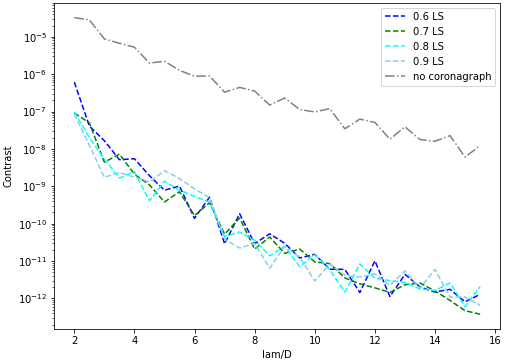}
    \includegraphics[width=.54\textwidth]{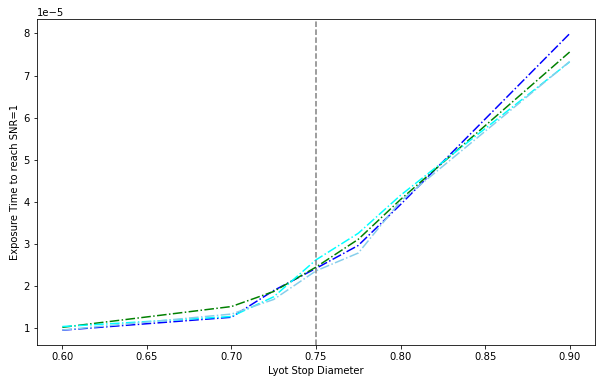}
    \caption{The design for the vector vortex coronagraph. Left: the contrast curves. For a vortex mask the only variation between design performance is due to random wavefront error, and the exposure time will be a deciding metric. Right: the exposure time to reach SNR$=1$ for a planet observation. The four curves represent the performance profiles as random wavefront error impact performance. For the vortex, due to the focal plane mask interfering light with itself, random wavefront error has a notable impact simulation to simulation. The optimized design (a Lyot Stop of 0.75) represents where the exposure time begins to reach a low within the factor of $\sim$2 introduced by the scatter of the random wavefront error.}
    \label{fig:vortex_design}
\end{figure}

Figures \ref{fig:lyot_design} and \ref{fig:vortex_design} show the optimization process for the two coronagraphs. Target designs are described in Table \ref{tab:coronagraph_designs}. Our target sizes will act as a starting point for integration, but we plan to test a range of Lyot stop sizes due to unanticipated sources of error in our testbed that we have yet to include in our model. With preliminary designs in mind, in the near future we will be ordering parts both externally and in-house and then testing them on SEAL.

   \begin{table}[ht]
\caption{\textbf{Preliminary coronagraph designs for the Vortex and Lyot.}}
 \label{tab:coronagraph_designs}
 \begin{center}       
 \begin{tabular}{|l|l|l|l}
 \hline
 \rule[-1ex]{0pt}{3.5ex}Vortex & Lyot Stop Diameter&  Topological Charge \\
   & (fractional) &  \\
\hline\hline
 & 0.75 & 4 \\
 \hline
 \rule[-1ex]{0pt}{3.5ex}Lyot & Lyot Stop Diameter &  Focal Plane Mask Diameter \\
    & (fractional) &  ($\lambda$/diameter) \\
 \hline\hline
& 0.6 & 3.0 \\
\hline

 \hline 
 \end{tabular}
 \end{center}
 \end{table}

\clearpage

\section{Real Time Control}
\label{sec: rtc}
The real time control system is based on the Keck RTC (\href{https://github.com/scetre/krtc}{KRTC}) project, originally developed for the Keck II infrared pyramid wavefront sensor. The KRTC uses some of the key CACAO\cite{2018SPIE10703E..1EG} functionalities such as shared memories and semaphores. So far, our focus has been oriented towards interfacing with the different SEAL hardware components. The following hardware is fully integrated with the KRTC solution:
\begin{enumerate}[noitemsep,topsep=0pt]
    \item ALPAO DM
    \item Boston Micromachines DM
    \item IRISAO segmented DM
    \item Andor Zyla Cameras
    \item EVT cameras
    \item FLIR Cameras
\end{enumerate}
Other hardware such as the Thorlabs Shack-Hartmann WFS and Meadowlark SLM are currently in the process of being made compatible with the KRTC.

The KRTC solution offers a python interface which is actively used in the Laboratory for Adaptive Optics (home to SEAL) to prototype and validate AO concepts. The next step of the development will be to transfer the python code into the KRTC library in C. The current library is limited to pyramid reconstruction in real time. The team is working to extend its capabilities to all of the techniques describe in this paper.

\section{Next Steps}
\label{sec:conclusion}
Currently, the refractive SEAL setup is broken up into several different subsystems in order to support the initial development of the different wavefront sensor arms and the characterization of the SLM. Our next step will be to integrate these subsystem into a single refractive SEAL setup that includes each of the major components shown in Figure \ref{fig:optical_layout} and demonstrate closed-loop woofer/tweeter control with the SHWFS and 3RPWFS. We will then aim to demonstrate our two second-stage wavefront sensors -- FAST and the Zernike WFS\cite{Doelman19} -- in conjunction with the SHWFS (FAST will correct residual atmospheric wavefront errors while the Zernike WFS will correct for primary mirror segment piston and tip/tilt errors). A key focus area for the SEAL testbed will also be the development of predictive control algorithms for ground-based high contrast imaging systems (see van Kooten et al.~2021\cite{vankooten21} in these proceedings).   


\acknowledgments 
This work was partially supported by the National Science Foundation AST-ATI Grant 2008822, the Heising-Simons Foundation, and the University of California Observatories. 
 
\bibliography{report}

\begin{thebibliography}{10}

\bibitem{ess2017}
``{Exoplanet Science Strategy},''  National Academies of Sciences, Engineering,
  and Medicine, National Academies Press, Washington, DC (2018).

\bibitem{2014PNAS..11112661M}
{Macintosh}, B., {Graham}, J.~R., {Ingraham}, P., {Konopacky}, Q., {Marois},
  C., {Perrin}, M., {Poyneer}, L., {Bauman}, B., {Barman}, T., {Burrows},
  A.~S., {Cardwell}, A., {Chilcote}, J., {De Rosa}, R.~J., {Dillon}, D.,
  {Doyon}, R., {Dunn}, J., {Erikson}, D., {Fitzgerald}, M.~P., {Gavel}, D.,
  {Goodsell}, S., {Hartung}, M., {Hibon}, P., {Kalas}, P., {Larkin}, J.,
  {Maire}, J., {Marchis}, F., {Marley}, M.~S., {McBride}, J.,
  {Millar-Blanchaer}, M., {Morzinski}, K., {Norton}, A., {Oppenheimer}, B.~R.,
  {Palmer}, D., {Patience}, J., {Pueyo}, L., {Rantakyro}, F., {Sadakuni}, N.,
  {Saddlemyer}, L., {Savransky}, D., {Serio}, A., {Soummer}, R.,
  {Sivaramakrishnan}, A., {Song}, I., {Thomas}, S., {Wallace}, J.~K.,
  {Wiktorowicz}, S., and {Wolff}, S., ``{First light of the Gemini Planet
  Imager},'' {\em Proceedings of the National Academy of Science}~{\bf 111},
  12661--12666 (Sept. 2014).

\bibitem{2019AA...631A.155B}
{Beuzit}, J.~L., {Vigan}, A., {Mouillet}, D., {Dohlen}, K., {Gratton}, R.,
  {Boccaletti}, A., {Sauvage}, J.~F., {Schmid}, H.~M., {Langlois}, M., {Petit},
  C., {Baruffolo}, A., {Feldt}, M., {Milli}, J., {Wahhaj}, Z., {Abe}, L.,
  {Anselmi}, U., {Antichi}, J., {Barette}, R., {Baudrand}, J., {Baudoz}, P.,
  {Bazzon}, A., {Bernardi}, P., {Blanchard}, P., {Brast}, R., {Bruno}, P.,
  {Buey}, T., {Carbillet}, M., {Carle}, M., {Cascone}, E., {Chapron}, F.,
  {Charton}, J., {Chauvin}, G., {Claudi}, R., {Costille}, A., {De Caprio}, V.,
  {de Boer}, J., {Delboulb{\'e}}, A., {Desidera}, S., {Dominik}, C., {Downing},
  M., {Dupuis}, O., {Fabron}, C., {Fantinel}, D., {Farisato}, G., {Feautrier},
  P., {Fedrigo}, E., {Fusco}, T., {Gigan}, P., {Ginski}, C., {Girard}, J.,
  {Giro}, E., {Gisler}, D., {Gluck}, L., {Gry}, C., {Henning}, T., {Hubin}, N.,
  {Hugot}, E., {Incorvaia}, S., {Jaquet}, M., {Kasper}, M., {Lagadec}, E.,
  {Lagrange}, A.~M., {Le Coroller}, H., {Le Mignant}, D., {Le Ruyet}, B.,
  {Lessio}, G., {Lizon}, J.~L., {Llored}, M., {Lundin}, L., {Madec}, F.,
  {Magnard}, Y., {Marteaud}, M., {Martinez}, P., {Maurel}, D., {M{\'e}nard},
  F., {Mesa}, D., {M{\"o}ller-Nilsson}, O., {Moulin}, T., {Moutou}, C.,
  {Orign{\'e}}, A., {Parisot}, J., {Pavlov}, A., {Perret}, D., {Pragt}, J.,
  {Puget}, P., {Rabou}, P., {Ramos}, J., {Reess}, J.~M., {Rigal}, F., {Rochat},
  S., {Roelfsema}, R., {Rousset}, G., {Roux}, A., {Saisse}, M., {Salasnich},
  B., {Santambrogio}, E., {Scuderi}, S., {Segransan}, D., {Sevin}, A.,
  {Siebenmorgen}, R., {Soenke}, C., {Stadler}, E., {Suarez}, M., {Tiph{\`e}ne},
  D., {Turatto}, M., {Udry}, S., {Vakili}, F., {Waters}, L.~B.~F.~M., {Weber},
  L., {Wildi}, F., {Zins}, G., and {Zurlo}, A., ``{SPHERE: the exoplanet imager
  for the Very Large Telescope},'' {\em A\&A}~{\bf 631},  A155 (Nov. 2019).

\bibitem{2016SPIE.9909E..0VB}
{Bailey}, V.~P., {Poyneer}, L.~A., {Macintosh}, B.~A., {Savransky}, D., {Wang},
  J.~J., {De Rosa}, R.~J., {Follette}, K.~B., {Ammons}, S.~M., {Hayward}, T.,
  {Ingraham}, P., {Maire}, J., {Palmer}, D.~W., {Perrin}, M.~D., {Rajan}, A.,
  {Rantakyr{\"o}}, F.~T., {Thomas}, S., and {V{\'e}ran}, J.-P., ``{Status and
  performance of the Gemini Planet Imager adaptive optics system},'' in [{\em
  Adaptive Optics Systems V}{\nolinebreak\hspace{0.1em}]},  {Marchetti}, E.,
  {Close}, L.~M., and {V{\'e}ran}, J.-P., eds., {\em Society of Photo-Optical
  Instrumentation Engineers (SPIE) Conference Series} {\bf 9909},  99090V (July
  2016).

\bibitem{2017arXiv171005417M}
{Milli}, J., {Mouillet}, D., {Fusco}, T., {Girard}, J.~H., {Masciadri}, E.,
  {Pena}, E., {Sauvage}, J.~F., {Reyes}, C., {Dohlen}, K., {Beuzit}, J.~L.,
  {Kasper}, M., {Sarazin}, M., and {Cantalloube}, F., ``{Performance of the
  extreme-AO instrument VLT/SPHERE and dependence on the atmospheric
  conditions},'' {\em Adaptive Optics for Extremely Large Telescopes 5,
  Conference Proceeding, Tenerife, Canary Islands, Spain} ,  arXiv:1710.05417
  (Oct. 2017).

\bibitem{vankooten21}
{van Kooten}, M. A.~M., {Jensen-Clem}, R., {Cetre}, S., {Ragland}, S., {Bond},
  C.~Z., {Fowler}, J., and {Wizinowich}, P., ``{Status of predictive wavefront
  control on Keck II adaptive optics bench: on-sky coronagraphic results},''
  {\bf 11823-43}, International Society for Optics and Photonics, SPIE,
  arXiv:2108.08932 (2021).

\bibitem{fast_seal}
Gerard, B.~L., Dillon, D., Cetre, S., Jensen-Clem, R., Yuzvinsky, T.~D., and
  Schmidt, H., ``{First Experimental Results of the Fast Atmospheric
  Self-coherent Camera Technique on the Santa cruz Extreme Adaptive optics
  Laboratory Testbed},''  {\bf 11823-76}, International Society for Optics and
  Photonics, SPIE (2021).

\bibitem{Doelman19}
Doelman, D.~S., Auer, F.~F., Escuti, M.~J., and Snik, F., ``Simultaneous phase
  and amplitude aberration sensing with a liquid-crystal vector-zernike phase
  mask,'' {\em Opt. Lett.}~{\bf 44},  17--20 (Jan 2019).

\bibitem{SLMManual}
{Meadowlark Optics Inc.}, {\em {1920 x 1152 XY Phase SLM. -- PCIe User
  Manual}}, {Rev. 1.0.3}~ed.

\bibitem{1976JOSA...66..207N}
{Noll}, R.~J., ``{Zernike polynomials and atmospheric turbulence.},'' {\em
  Journal of the Optical Society of America (1917-1983)}~{\bf 66},  207--211
  (Mar. 1976).

\bibitem{Neyman04}
{Neyman}, C., ``{Atmospheric Parameters for Mauna Kea},'' tech. rep. (2004).

\bibitem{sanchez_design_2020}
Sanchez, D.~F., Chun, M.~R., Bond, C.~Z., and Hinz, P.~M., ``Design study for a
  3-sided reflective pyramid wavefront sensor for {Shane} {AO},'' in [{\em
  Adaptive {Optics} {Systems} {VII}}{\nolinebreak\hspace{0.1em}]},  Schmidt,
  D., Schreiber, L., and Vernet, E., eds.,  200, SPIE, Online Only, United
  States (Dec. 2020).

\bibitem{por2018hcipy}
Por, E.~H., Haffert, S.~Y., Radhakrishnan, V.~M., Doelman, D.~S., Van~Kooten,
  M., and Bos, S.~P., ``{High Contrast Imaging for Python (HCIPy): an
  open-source adaptive optics and coronagraph simulator},'' in [{\em Adaptive
  Optics Systems VI}{\nolinebreak\hspace{0.1em}]},  {\em Proc. {{SPIE}}} {\bf
  10703} (2018).

\bibitem{scc_orig}
{Baudoz}, P., {Boccaletti}, A., {Baudrand}, J., and {Rouan}, D., ``{The
  Self-Coherent Camera: a new tool for planet detection},'' in [{\em IAU
  Colloq. 200: Direct Imaging of Exoplanets: Science \&
  Techniques}{\nolinebreak\hspace{0.1em}]},  {Aime}, C. and {Vakili}, F., eds.,
   553--558 (Jan. 2006).

\bibitem{Sivaramakrishnan}
{Sivaramakrishnan}, A., {Koresko}, C.~D., {Makidon}, R.~B., {Berkefeld}, T.,
  and {Kuchner}, M.~J., ``{Ground-based Coronagraphy with High-order Adaptive
  Optics},'' {\em ApJ}~{\bf 552},  397--408 (May 2001).

\bibitem{Makidon}
{Makidon}, R.~B., {Sivaramakrishnan}, A., {Koresko}, C.~D., {Berkefeld}, T.,
  {Kuchner}, M.~J., and {Winsor}, R.~S., ``{Ground-based coronagraphy with
  high-order adaptive optics},'' in [{\em Adaptive Optical Systems
  Technology}{\nolinebreak\hspace{0.1em}]},  {Wizinowich}, P.~L., ed., {\em
  Society of Photo-Optical Instrumentation Engineers (SPIE) Conference Series}
  {\bf 4007},  989--998 (July 2000).

\bibitem{Mawet2009}
{Mawet}, D., {Trauger}, J.~T., {Serabyn}, E., {Moody}, Dwight~C., J., {Liewer},
  K.~M., {Krist}, J.~E., {Shemo}, D.~M., and {O'Brien}, N.~A., ``{Vector vortex
  coronagraph: first results in the visible},'' in [{\em Techniques and
  Instrumentation for Detection of Exoplanets IV}{\nolinebreak\hspace{0.1em}]},
   {Shaklan}, S.~B., ed., {\em Society of Photo-Optical Instrumentation
  Engineers (SPIE) Conference Series} {\bf 7440},  74400X (Aug. 2009).

\bibitem{Ruane2018}
{Ruane}, G., {Riggs}, A., {Mazoyer}, J., {Por}, E.~H., {N'Diaye}, M., {Huby},
  E., {Baudoz}, P., {Galicher}, R., {Douglas}, E., {Knight}, J., {Carlomagno},
  B., {Fogarty}, K., {Pueyo}, L., {Zimmerman}, N., {Absil}, O., {Beaulieu}, M.,
  {Cady}, E., {Carlotti}, A., {Doelman}, D., {Guyon}, O., {Haffert}, S.,
  {Jewell}, J., {Jovanovic}, N., {Keller}, C., {Kenworthy}, M.~A., {Kuhn}, J.,
  {Miller}, K., {Sirbu}, D., {Snik}, F., {Wallace}, J.~K., {Wilby}, M., and
  {Ygouf}, M., ``{Review of high-contrast imaging systems for current and
  future ground- and space-based telescopes I: coronagraph design methods and
  optical performance metrics},'' in [{\em Space Telescopes and Instrumentation
  2018: Optical, Infrared, and Millimeter Wave}{\nolinebreak\hspace{0.1em}]},
  {Lystrup}, M., {MacEwen}, H.~A., {Fazio}, G.~G., {Batalha}, N., {Siegler},
  N., and {Tong}, E.~C., eds., {\em Society of Photo-Optical Instrumentation
  Engineers (SPIE) Conference Series} {\bf 10698},  106982S (Aug. 2018).

\bibitem{2018SPIE10703E..1EG}
{Guyon}, O., {Sevin}, A., {Gratadour}, D., {Bernard}, J., {Ltaief}, H.,
  {Sukkari}, D., {Cetre}, S., {Skaf}, N., {Lozi}, J., {Martinache}, F.,
  {Clergeon}, C., {Norris}, B., {Wong}, A., and {Males}, J., ``{The compute and
  control for adaptive optics (CACAO) real-time control software package},'' in
  [{\em Adaptive Optics Systems VI}{\nolinebreak\hspace{0.1em}]},  {Close},
  L.~M., {Schreiber}, L., and {Schmidt}, D., eds., {\em Society of
  Photo-Optical Instrumentation Engineers (SPIE) Conference Series} {\bf
  10703},  107031E (July 2018).

\end{thebibliography}

\bibliographystyle{spiebib} 


\end{document}